\documentclass[preprint]{aastex631}

\shorttitle{AASTeX v6.3.1 Sample article}
\shortauthors{Penza et al.}
\graphicspath{{./}{figures/}}
\begin{document}

\title{
%Forecasting plage and spot coverages for Solar Cycle 25
Prediction of sunspot and plage coverage for Solar Cycle 25
}

\correspondingauthor{Serena Criscuoli}
\email{scriscuo@nso.edu}

\author[0000-0002-3948-2268]{Valentina Penza}
\affiliation{Dipartimento di Fisica, Universit\`a di Roma Tor Vergata, Via della Ricerca Scientifica 1, Roma, 00133, Italy}

\author[0000-0002-2276-3733]{Francesco Berrilli}
\affiliation{Dipartimento di Fisica, Universit\`a di Roma Tor Vergata, Via della Ricerca Scientifica 1, Roma, 00133, Italy}

\author[0000-0002-1155-7141]{Luca Bertello}
\affil{National Solar Observatory, 3665 Discovery Dr., Boulder, CO 80303, USA}

\author[0000-0003-4898-2683]{Matteo Cantoresi}
\affiliation{Dipartimento di Fisica, Universit\`a di Roma Tor Vergata, Via della Ricerca Scientifica 1, Roma, 00133, Italy}

\author[0000-0002-4525-9038]{Serena Criscuoli}
\affil{National Solar Observatory, 3665 Discovery Dr., Boulder, CO 80303, USA}



\begin{abstract}

Solar variability occurs over a broad range of spatial and temporal scales,
from the Sun's brightening over its lifetime to the fluctuations commonly associated with 
magnetic activity over minutes to years. 
The latter activity includes most prominently the 11-year sunspot solar cycle 
and its modulations.  Space weather events, in the form of
solar flares, solar energetic particles, coronal mass ejections, and geomagnetic storms, 
have long been known to approximately follow the solar cycle occurring more frequently 
at solar maximum than solar minimum. These events
can significantly impact our advanced technologies and critical infrastructures,
making the prediction for the strength of future solar cycles particularly important. 

Several methods have been proposed to predict the strength of the next solar cycle, cycle 25, 
with results that are generally not always consistent. Most of these methods are based
on the international sunspot number time series, or other indicators of
solar activity. We present here a new approach that
uses more than 100 years of measured fractional areas of the visible solar disk covered 
by sunspots and plages 
and an empirical relationship for each of these two indices of solar activity in even-odd cycles. 
We anticipate that cycle 25 will peak in 2024 and will last for about 12 years, slightly
longer than cycle 24. We also found that, 
in terms of sunspot and plage areas coverage, the amplitude of cycle 25 will be substantially similar or slightly higher than cycle 24.

\end{abstract}



\section{Introduction} \label{sec:intro}

The variability of the Sun's magnetic activity is the main forcing that drives change in the near-Earth space environment, the heliosphere, and the interplanetary medium. This variability is
characterized by a large number of phenomena that act over a wide range of temporal scales, from minutes to centuries. A particularly important class of short-lived phenomena, up to a few days, includes events
such as solar flares and coronal mass ejections that are responsible for the space weather. These events have a significant impact on the functionality of satellites, communication networks, electric power grid and technological infrastructures
%instruments
\citep[e.g.][]{schwenn2006,Plainaki2020}. Instead, solar activity over longer time-scales modifies the so-called space climate \citep{versteegh2005} that contributes to determine the conditions of the Earth's upper and lower atmosphere and
%, to a lesser extent,  
the Earth's climate \citep[e.g.][]{Bordi2015,matthes2017,Bigazzi2020,Lockwood2020} and the cosmic ray fluxes in the interplanetary space and at Earth \citep[e.g.][]{usoskin2002,Berrilli2014,Fiandrini2021}. Similarly, stellar irradiance and its variations affect exoplanets' atmospheres and their habitability \citep[e.g.][]{meadows2018,Galuzzo2021}.
The most evident modulation of the large-scale solar magnetic field is its 11-year cycle, during which the Sun increases and weakens its magnetic activity, associated with a reversal of the dominant polarities in the
polar regions.
%it increases and decreases and at the end it is subject to a reverse of polarity. 
This cycle is accompanied by variations in phase of the appearance
of magnetic structures on the solar surface, such as sunspots and plages. The dark sunspots produce a luminosity defect, while the bright plages overcompensate with an excess \citep[e.g.][]{Foukal1988}. The final result is that the total solar irradiance varies by about $ 0.1 \% $, in phase with the magnetic activity
%field
\citep[e.g.][]{wilson1978, hudson1988,kopp2016}. The strength of the different cycles is not constant, and can vary quite significantly as indicated by the presence of grand minima periods \citep[e.g.][]{Vecchio2017} such as the Maunder minimum during the years of 1645 to 1715 \citep[e.g.][]{hathaway2015}.\\
The capability to predict the behavior of the solar activity has become of paramount relevance,
%fundamental importance
given its enormous impact on human activities on Earth and in space. 
Providing a detailed description of the literature regarding the prediction of the next 25\textsuperscript{th} solar cycle (hereafter SC25) is beyond the scope of this paper. We therefore refer to the review of solar cycle prediction methods, with particular focus on forecasts for SC25, given in \cite{petrovay2020}. The various forecasts differ in the adopted methodology (e.g., surface flux transport models, identification of particular “termination” events, Shannon entropy estimates, machine learning regression methods, etc.) and in the physical observable on which the prediction is based. In particular, the methods typically fall into three categories: precursor, model-based and extrapolation methods. For the prediction of SC25 different studies have been published that use the sunspot number \citep[e.g.][]{McIntosh2020,singh21}, the geomagnetic activity indices \citep[e.g.][]{singh21}, the flare number \citep[e.g.][]{janssens21} and the solar magnetic flux or dipole momentum \citep[e.g.][]{Cameron2016,bhowmik18,upton2018,labonville2019}.\\
The aim of this work is to estimate the coverage of sunspot and plage areas during the next SC25. We use an appropriate functional form for solar cycles derived from the correlation between cycles for the period 1874-2019. This approach has two important aspects: First, it is based on the observed characteristics of the cycles and does not use sophisticated physical models which unfortunately suffer from the difficulties connected to the complex physical processes involved and to the inherent dynamical complexity of the solar cycle \citep[e.g.][]{Bushby2004,Consolini2009,Charbonneau2020};
Second, the coverage in the area of sunspots and plages
%In this work, we present a forecast for the sunspot and plage coverages which 
can be used as proxies to predict other activity indices important for space-climate, including spectral and total solar irradiance variability %and are a fundamental ingredient for the reconstruction of the spectral and total solar irradiance variability 
\citep[e.g.][]{Foukal1988,berrilli2020,Petrie2021}.

\section{Active Regions Parametrization} \label{sec:1}

We characterize the shape of each solar cycle
through a unique functional form. Among those proposed in literature  \citep[e.g.][]{Baranov2008,hathaway94}, we choose the following %one-parameter 
parametric fit suggested in \cite{Volobuev}:
%%%%%%%%%%%%%%%%%%%%%
\begin{equation}
\label{cycle_form}
x_{k}(t) =  \left(\frac{t - T0_{k}}{Ts_{k}}\right)^{2} e^{-\left(\frac{t - T0_{k}}{Td_{k}}\right)^{2}} \quad \quad \textrm{for} \quad T0_{k} < t < T0_{k} + \tau_{k}
\end{equation}
%%%%%%%%%%%%%%%%%%%%%%%
where $T0_{k}$ is the initial time of cycle $k$ \citep[published in][]{hathaway94, hathaway2015} while $Ts_{k}$ and $Td_{k}$ are two free parameters. As discussed in \cite{Volobuev}, there is a strong linear correlation between these two parameters, which reduces the fit to a single parameter ($Ts_{k}$). This is not surprising, because the $Ts_{k}$ parameter is related to the time of the rising phase, while the $Td_{k}$ value determines the cycle amplitude. The relation between these two quantities is known and reported in literature: cycles with large amplitude ($Ts_{k}$ smaller) present a shorter time of rising to maximum (shorter $Td_{k}$). This is known as Waldmeier effect \citep[e.g.][]{hathaway94,hazra2015}.\\ 
%%%%%%%%%%%%%%%%%%%%%%%%% 
\begin{figure}[htbp]
\centering
\includegraphics[width=0.47\linewidth,trim=4cm 0cm 4cm
0cm,clip]{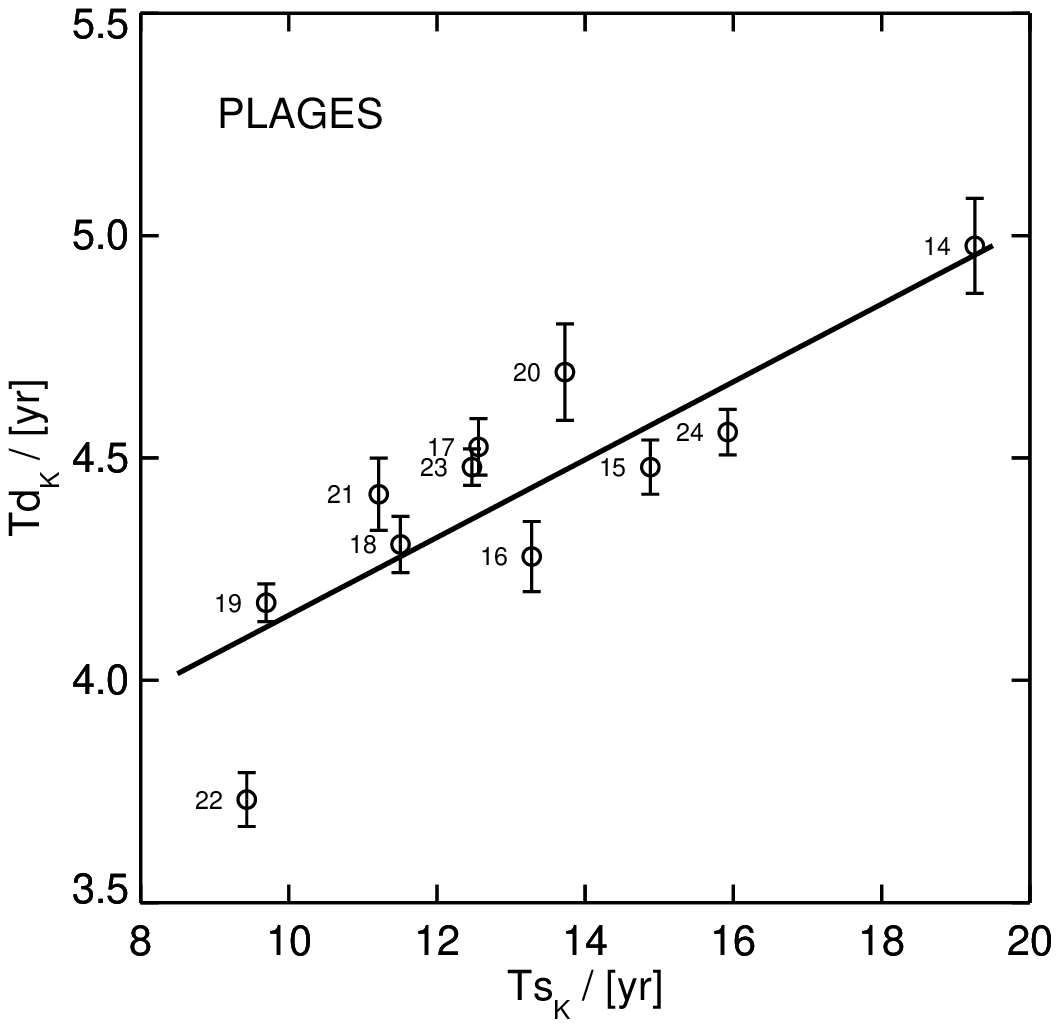}
\includegraphics[width=0.47\linewidth,trim=4cm 0cm 4cm 0cm,clip]{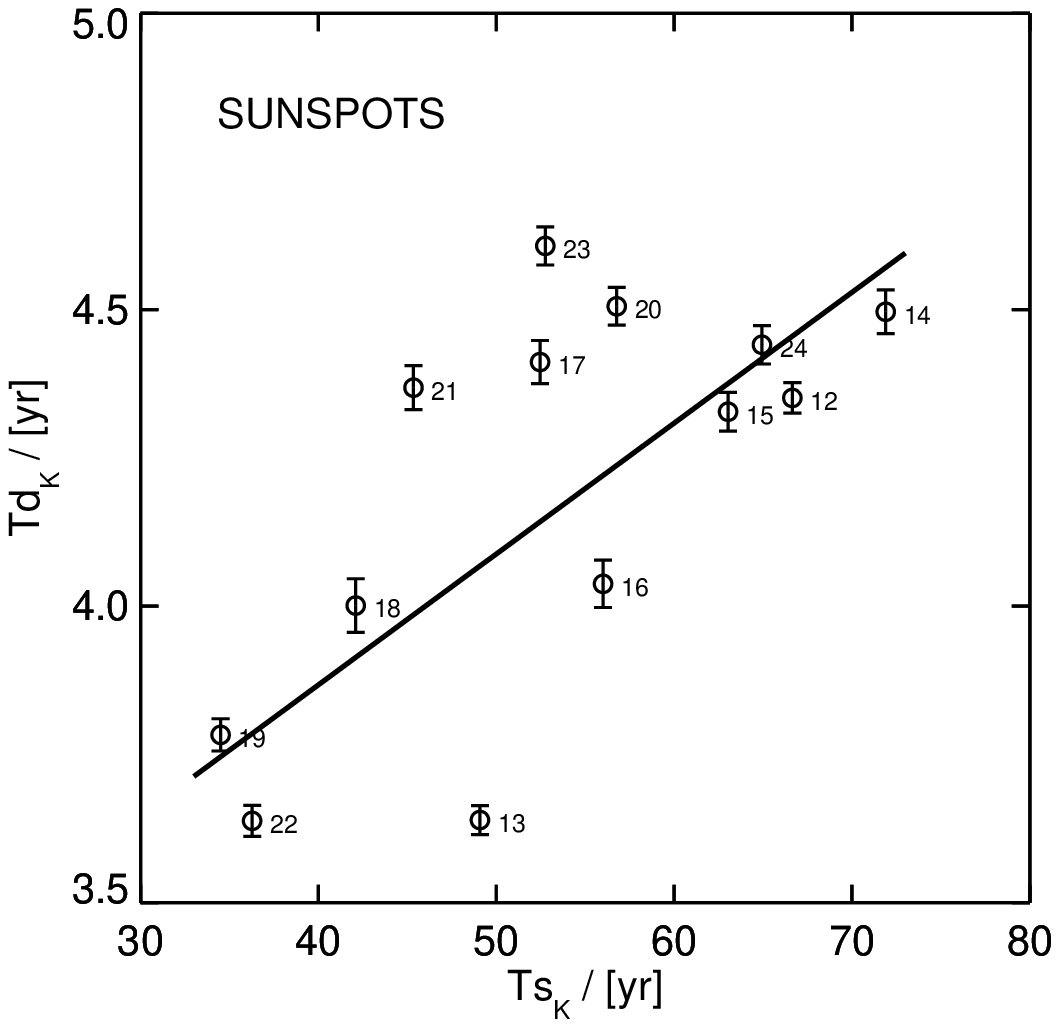}
\caption{
Solar cycle-to-cycle relationship between the parameters $Td_k$ and 
$Ts_k$ for plage (left) and sunspot
(right) areas. Two additional solar cycle numbers, 12 and 13, have been included in the
analysis of sunspot areas. The open circles with 1-$\sigma$ error bars show the data 
averaged over individual cycles. The regression lines (continuous lines) are given by Eq. 2.
}
\label{TsvsTd_plot}
\end{figure}
%%%%%%%%%%%%%%%%%%%%%%%%%%%%%%%%
%%%%%%%%%%%%%%%%%%%%%%%%% 
\begin{figure}[h]
\centering
\includegraphics[height=12 cm ,width=18 cm]{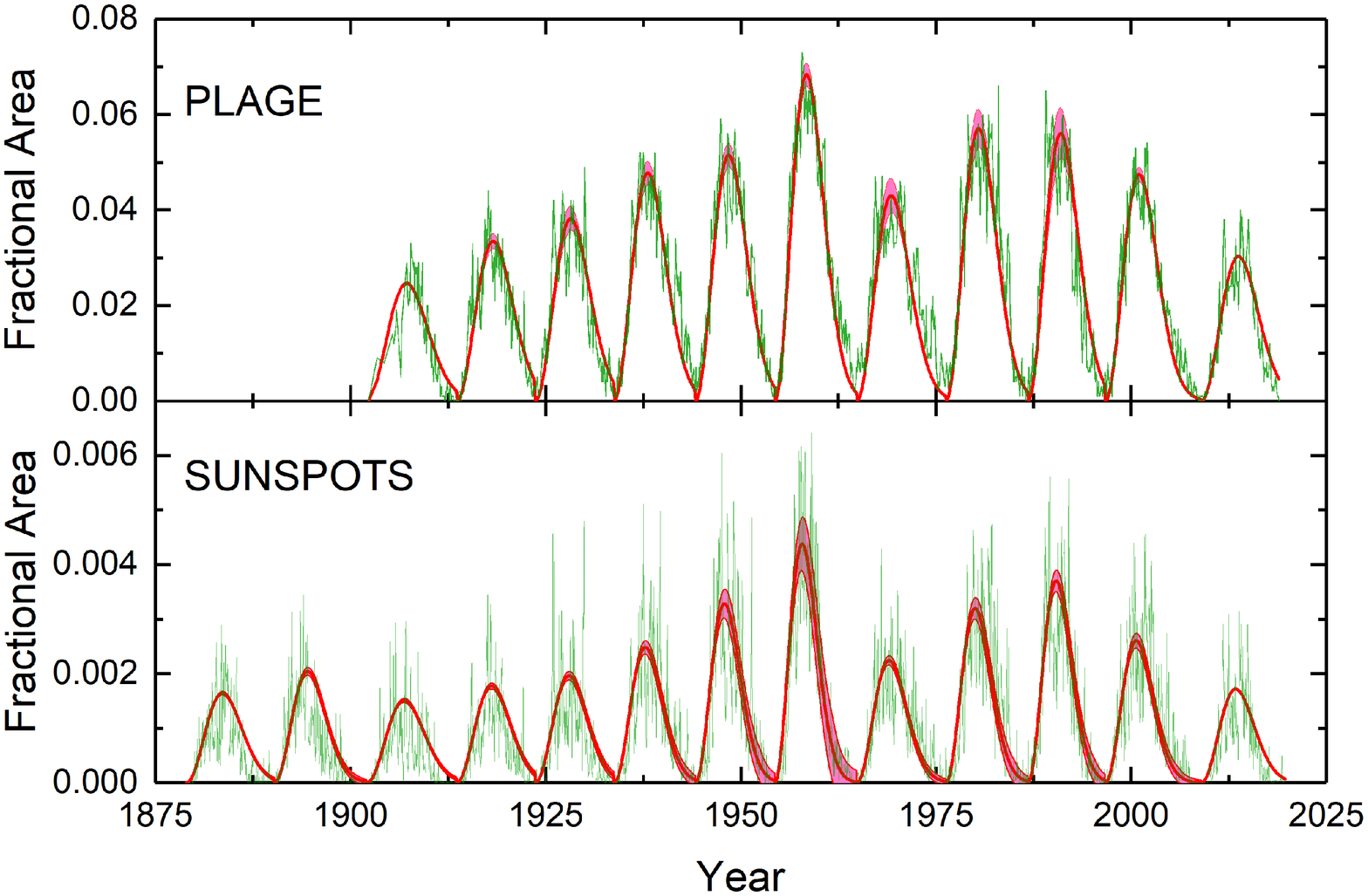}
\caption{\textit{Upper panel:} %Parametric reconstruction of plage coverage (blue) with confidence range (gray). For comparison the plage composite from \cite{chatzistergos19} (red). \textit{Lower panel:} The parametric reconstruction of Sunspot coverage (blue) with confidence range (gray). For comparison the Sunspot composite from \cite{mandal2020} (red).}.
reconstructed plage area (red solid line) with confidence range (light red pattern) and observed plage composite (green solid line) by \cite{chatzistergos19}.
\textit{Lower panel:} reconstructed Sunspot area (red solid line) with confidence range (light red pattern) and observed sunspot area  composite (green solid line) from \cite{mandal2020}. It should be noted that our method has been developed in order to provide an estimate of the coverage of plage and sunspot on an annual scale and therefore cannot reproduce shorter time scale variability (e.g. months). This explains the difference between the observed plage and sunspot coverages (green solid lines) and the reconstructed ones (red solid line). The ability to capture long-term behavior by the functional form is also evident.
}
\label{AR_parametric}
\end{figure}
%%%%%%%%%%%%%%%%%%%%%%%%%%%%%%%%
%We also have used 
We adopt Eq.~\ref{cycle_form} to fit, cycle by cycle, the plage and sunspots coverage data. For this purpose
%In particular, 
we used two data sets (composites) available from the Max Plank Institute (MPI) website (http://www2.mps.mpg.de/projects/sun-climate/data.html). The first composite consists of plage area derived from Ca II K spectroheliogram observations covering the period 1892 to 2019 \citep{chatzistergos19}. The second composite includes measurements of daily total sunspot area for the period 1874-2019, calculated after cross-calibration of measurements by different observers \citep{mandal2020}. 
The relation between the $Ts_{k}$ and $Td_{k}$ for plage and sunspot areas is shown in Fig.~\ref{TsvsTd_plot}. A linear fit to these data produces:
%By the fits, we obtained the following relations between $Ts_{k}$ and $Td_{k}$:
%%%%%%%%%%%%%%%%%%%%%%%%
\begin{eqnarray}
\label{TsvsTd}
\nonumber Td_{k}^{plage} = (0.09 \pm 0.01) Ts_{k}^{plage} + (3.27 \pm 0.10)~~~ yr  \\
Td_{k}^{spot} = (0.022 \pm 0.001)Ts_{k}^{spot} + (2.98 \pm 0.04) ~~~  yr . 
\end{eqnarray}
%%%%%%%%%%%%%%%%%%%%%%%%%
%The correlations corresponding to eqs.~\ref{TsvsTd} are shown in Fig.~\ref{TsvsTd_plot}; 
%The errors are relative to the uncertainties calculated on the %fit parameters \citep{bevington}.
The Pearson correlation coefficients are r=0.81 and r=0.72 for plage and spot fit, respectively. A t-test was performed to determine the statistical significance of the computed correlation coefficients. We found there is a non-zero correlation between $Ts_{k}$ and $Td_{k}$, at a confidence level greater than 99\% for the plages and greater than 95\% for the sunspots.\\
By inserting these relationships in Eq.~\ref{cycle_form}, we obtain a one-parameter functional form for the shape of the cycles. We repeat the fits, cycle by cycle, and we obtain two dataset of $Ts_{k}$ values.\\
The active region coverages ($A(t)_{plage}$ and $A(t)_{spot}$) for the entire analyzed period are thus reproduced as:
%%%%%%%%%%%%%%%%%%%%%
\begin{equation}
\label{cycle_rec}
A(t) =  \sum_{k} x_{k}(t) 
\end{equation}
%%%%%%%%%%%%%%%%%%%%%%%
where the cycle number k goes from the 14-th (begin data = 1902 January) to the 23-th (begin data = 1996 August) for plage data and from the 12-th (begin data = 1878 December) for sunspot data. 
The parametric reconstructions for plage and sunspot area are shown in Fig.~\ref{AR_parametric}. A confidence region is estimated by taking into account the errors on the fit parameters.\\ 
%This region is more evident in the plage case, while it is practically indistinguishable in the sunspots case. \\
%The parameterization carried out in the previous paragraph allows %to characterize each cycle through a single parameter ($Ts_{k}$). 

\section{Relation between even and odd cycles} \label{sec:2}

\begin{figure}[htbp]
\centering
\includegraphics[width=0.47\linewidth,trim=4cm 0cm 4cm
0cm,clip]{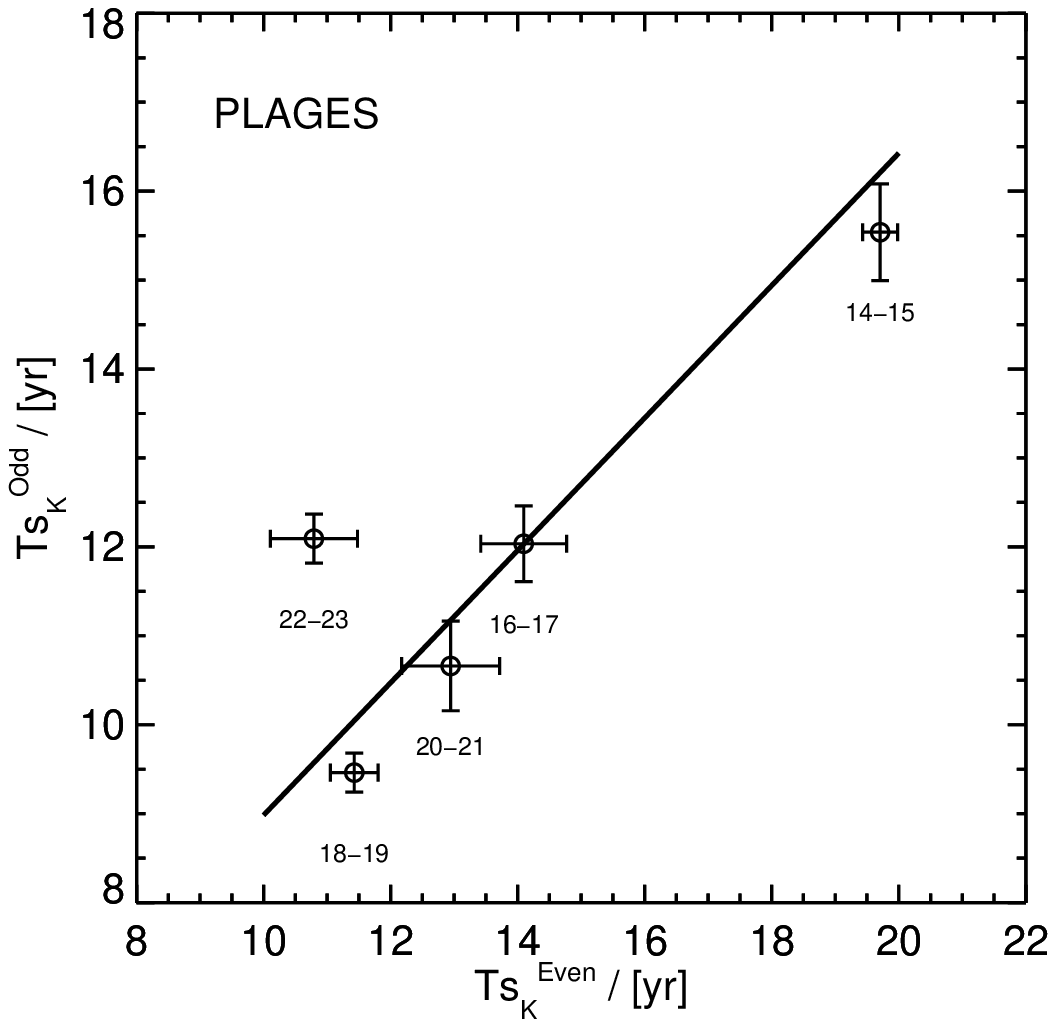}
\includegraphics[width=0.47\linewidth,trim=4cm 0cm 4cm 0cm,clip]{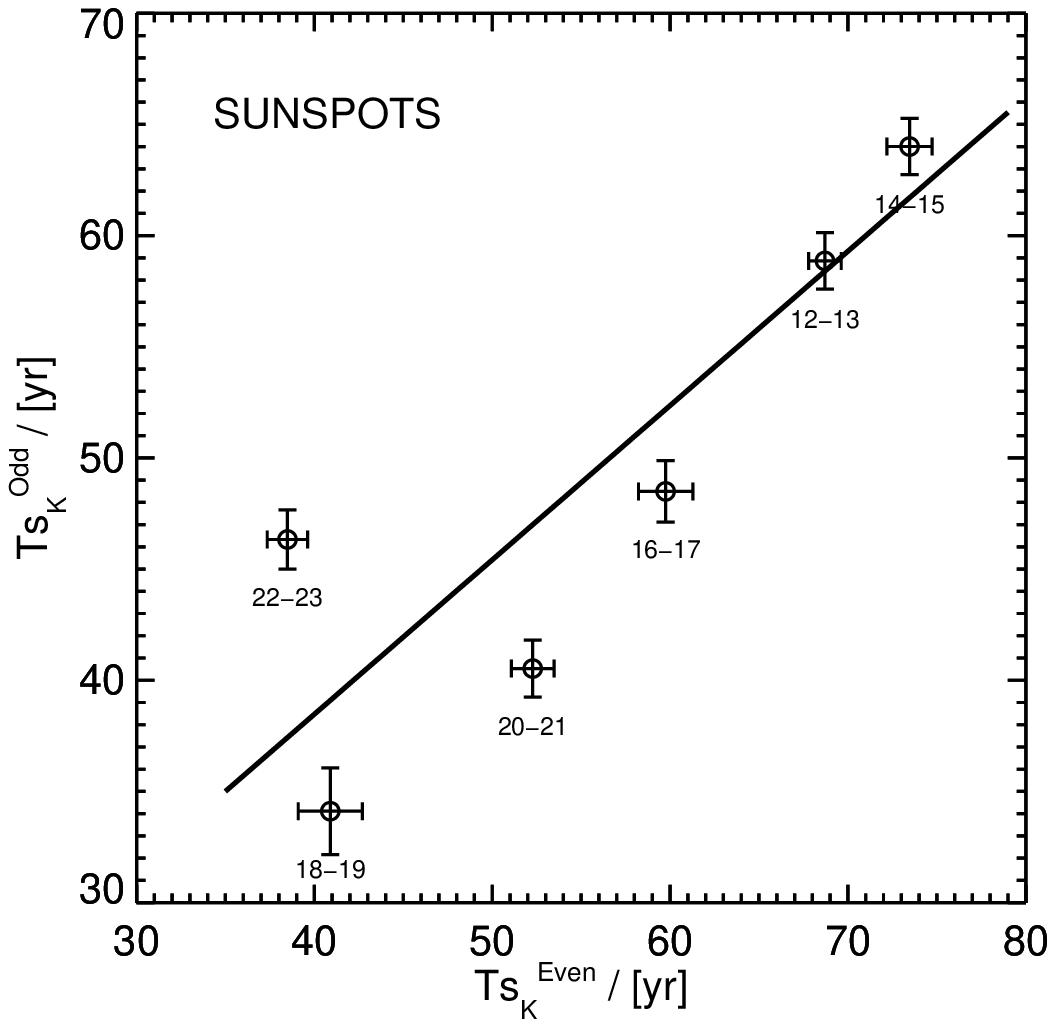}
\caption{
Even-odd cycle relationship between the parameters $Ts_k$ for plage (left) and sunspot (right) areas.  The open circles with 1-$\sigma$ error bars in both directions show the cycle-averaged pair of data. 
The regression lines (continuous lines), given by Eq. \ref{even_odd},  were computed using the procedure described in \citet{PresTeukVettFlan92}.
}
\label{Ts_even_odd}
\end{figure}

Through the fits of the observed active region coverages, we have obtained the single parameter characterizing the last eleven cycles for the plages and the last thirteen cycles for the sunspots.\\
We couple these values in pairs, i.e. as even-odd cycles, taking into account that the complete magnetic cycle of the Sun is composed of two consecutive cycles. By plotting the even-odd parameters, one versus the other, we obtain the correlations shown in Fig. \ref{Ts_even_odd}, whose fits are as follows

%%%%%%%%%%%%%%%%%%%%%%%%
\begin{eqnarray}
\label{even_odd}
\nonumber Ts_{o}^{plage} = (0.74 \pm 0.08) Ts_{e}^{plage} + (1.5 \pm 1.1)~~~ yr  \\
Ts_{o}^{spot} = (0.69 \pm 0.05) Ts_{e}^{spot} + (11 \pm 3) ~~~  yr .      
\end{eqnarray}
%%%%%%%%%%%%%%%%%%%%%%%%%

where the subscript "e" means "even" and "o" means "odd".
 The Pearson correlation coefficient is 0.86 for both fits, at a confidence level greater than 90\% and 95\% for the plages and sunspots respectively.
We highlight the fact that by coupling the parameters% of the even-odd cycles differently 
%%%%%%%%%%%%%%%%%%%%%%%%% 
%%%%%%%%%%%%%%%%%%%%%%%%%%%%%%%%
%%%%%%%%%%%%%%%%%%%%%%%%% 
\begin{figure}[h]
\centering
\includegraphics[height=10 cm ,width=15 cm]{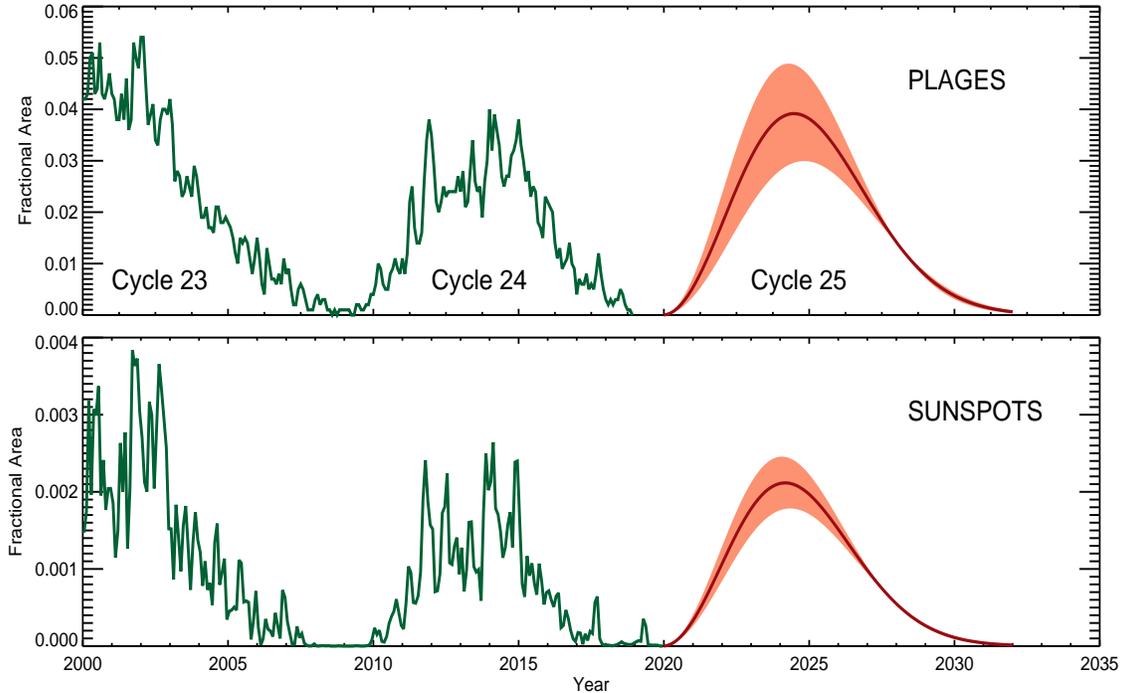}
\caption{
Top: Observed monthly plage coverage (green curve) during cycles 23 and 24, and prediction
for cycle 25 (red curve).
The shadow red area defines the lower and upper limits for the prediction.
Bottom: Same, but for sunspot coverage.}
\label{predict}
\end{figure}
%%%%%%%%%%%%%%%%%%%%%%%%%%%%%%%%
in an odd-even manner (e.g. Cycle 15 with Cycle 16) the correlation is completely lost. This is in agreement with the well known empirically derived Gnevyshev-Ohl rule \citep{gnevyshev}, which states that the strength of an even cycle is lower than the strength of the subsequent odd cycle.\\
By inserting the $Ts_{24}$ values derived from sunspot and plage observations and $T0_{25}$ corresponding to December 2019 in Eq.~\ref{even_odd}, we are able to estimate the $Ts_{25}$ values and therefore the shape of the variation of sunspot and plage areas over SC25 using Eq.~\ref{cycle_form}.
%We can apply Eq. \ref{even_odd} by inserting as $Ts_{e}$ the data provided by the fit on SC24, so we obtain the value for the two parameters $Ts_{o}$  of plages and sunspots for SC25: $Ts_{25}^{plage} = 13.8 \pm 2.5$ yr and $Ts_{25}^{spot} = 56 \pm 6$ yr, By inserting these values in Eq.~\ref{cycle_form}, we are able to estimate the %corresponding curves for the Cycle 25, that are shown in Fig.\ref{predict}.
Results are illustrated in Fig.~\ref{predict}. The confidence levels shown in the figure are the envelopes of the curves computed by varying the $T_{s_{25}}$ and $T_{d_{25}}$ values within their 1 $\sigma$ confidence level in Eq.~\ref{cycle_form}. The plot shows that the maximum of plage area will occur in April 2024, with an uncertainty of three months, and that the maximum sunspot area coverage will occur in January 2024, with an uncertainty of two months. The plots also show that, within the confidence level, the amplitudes of the area coverage of sunspot and plage are similar or slightly higher  to those observed during cycle 24. 
%In order to estimate the confidence interval for the amplitude of SC25, we inserted Eq. \ref{Ts_even_odd} in Eq. \ref{TsvsTd_plot} and substitute in Eq. \ref{cycle_form} and we did vary the coefficients estimated by the fit inside their $1 \sigma$ interval. In the band plotted in Fig.~\ref{predict} it is reported the envelope of the curves obtained from Eq.~\ref{cycle_form} following this procedure. %band in which all the curves, obtained from Eq.~\ref{cycle_form} with this procedure, appear.

\section{Conclusions} \label{sec:3}
We have presented a prediction of the area coverage of sunspot and plages during solar cycle 25. Our method is based on the empirical  correlation between the shapes (amplitude and duration) of even and odd cycles derived from the analysis of more than 100 years of observations. Similar relations between properties of subsequent cycles, in particular the even-odd cycles relation,  have been empirically derived by using solar sunspot number and other activity proxies in numerous studies \citep[e.g.][]{mursula2001,gupta2007,tlatov2013, takalo2020, takalo2021}. 
Although some authors suggested that the even-odd relation might be the signature of a relic magnetic field \citep[][and references therein]{mursula2001}, the physical mechanisms responsible for the even-odd relation are not clear \citep{hathaway2015} and it is worth noticing that the relation has been questioned by some authors \citep{zolotova2015}.

The plage area maximum is predicted for April 2024, with an uncertainty of about three months, instead the sunspot area maximum is predicted for January 2024, with an uncertainty of about two months. By combining these two results and by assuming a perfect phase between the two quantities, we can hypothesize the maximum intensity level of the SC25 will be in February-March 2024.
By assuming that the sunspot and plage coverage are indicative of the cycle intensity, our prediction is a level of solar activity for the SC25 similar or slightly higher to that of SC24. This is in agreement with  most of the forecasts presented in the literature or using the first available values of the smoothed Solar Sunspot Number \citep[e.g.][]{bhowmik18,petrovay2020,Carrasco_2021} and with the conclusions of the international NOAA/NASA co-chaired SC25 Prediction Panel.

%However, our prediction differs from others  because it 
We want to stress here that our approach is, to our knowledge, the only one that provides a forecast for the coverage of plage and sunspot areas. These quantities are proxies of the solar magnetic activity, can be used to derive other activity indices important for space-climate (e.g. \ion{Mg}{2}, F10.7, flare indices, etc.)  and are fundamental ingredients to estimate both total and spectral irradiance variability.\\
%\textbf{The presented method is based on the derived empirical  correlation between the shapes (amplitude and duration) of even and odd cycles}. Similar relations between properties of subsequent cycles, in particular the even-odd cycles relation,  have been empirically derived by using solar sunspot number and other activity proxies in numerous studies \citep[e.g.][]{mursula2001,tlatov2013, takalo2020, takalo2021}. 
%Although some authors suggested that the even-odd relation might be the signature of a relic magnetic field \citep[][and references therein]{mursula2001}, the physical mechanisms responsible for the even-odd relation are not clear \citep{hathaway2015} and it is worth noticing that the relation has been questioned by some authors \citep{zolotova2015}.\\\
%

\begin{acknowledgments}
The authors are grateful to Dr. Lisa Upton for providing insightful comments on an early version of the manuscript. The National Solar Observatory is operated by the Association of Universities for Research in Astronomy, Inc. (AURA), under cooperative agreement with the National Science Foundation. M. Cantoresi is supported by the Joint Research PhD Program in “Astronomy, Astrophysics and Space Science” between the universities of Roma Tor Vergata an Roma Sapienza, and INAF.
\end{acknowledgments}

%\bibliography{cycle25}{}
\bibliography{cycle25}{}
\bibliographystyle{aasjournal}



\end{document}